\documentclass{appolb}
\usepackage{epsfig}
\usepackage{amssymb}

\newcommand{\beq}{\begin{eqnarray}}
\newcommand{\eeq}{\end{eqnarray}}
\newcommand{\nneeq}{\nonumber \end{eqnarray}}

\newcommand{\nn}{\nonumber \\}
\newcommand{\es}{& = &}

\newcommand{\ps}{& + &}

\newcommand{\ts}{& \times &}
\newcommand{\nt}{\nn \ts}
\newcommand{\np}{\nn \ps}

\newcommand{\cM}{ {\cal M} }
\newcommand{\cH}{ {\cal H} }

\newcommand{\cG}{ {\cal G} }

\newcommand{\cU}{ {\cal U} }

\begin{document}
\title{    Perturbative formulae for relativistic interactions of effective particles }
\author{   Stanis{\l}aw D. G{\l}azek 
\address{  Institute of Theoretical Physics,
           Faculty of Physics, 
           University of Warsaw                     }}
\date{     20 April, 2012}
\maketitle
\begin{abstract}
The concept of effective particles as degrees of
freedom in a relativistic quantum field theory is
defined using a non-perturbative renormalization
group procedure for Hamiltonians. However, every
candidate for a basic physical theory appears to
require an initial perturbative search for the set
of interaction terms that may provide a basis with
which the full effective theory Hamiltonian could
be constructed in a series of successive 
approximations. This article describes the
required perturbative expansion and illustrates it 
with a set of general 4th-order formulae. 
\end{abstract}
%\pacs{ 11.10.Gh, 12.38.-t, 12.39.-x }

\vskip-5in
\hfill {\bf IFT/12/01}
\vskip5.9in

%%%%%%%%%%%%%%%%%%%%%%%%
\section{ Introduction }
%%%%%%%%%%%%%%%%%%%%%%%%

The construction of relativistic interactions of
effective particles that is described in this
article is designed for a quantum theory in which
one can introduce a front form (FF) of Hamiltonian
dynamics~\cite{DiracFF}. The construction draws on
the previous work on similarity renormalization
group (SRG) procedure for Hamiltonians
\cite{GlazekWilson1}, flow equation \cite{Wegner1}, 
and renormalization group procedure for effective 
particles (RGPEP)~\cite{RGPEP}. It is the latter 
that is used here.

The key limitation of the RGPEP in comparison to
the general SRG procedure~\cite{GlazekWilson1} is
that in RGPEP one focuses on coefficients of
operators in an operator basis built from products
of creation and annihilation operators for
effective particles\footnote{For brevity, the
creation and annihilation operators will be
sometimes commonly called particle operators.}
rather than on the Hamiltonian matrix elements.
The requirement that the effective dynamics is
obtained by universally rotating the bare particle
operators to corresponding effective ones imposes
unitary constraints on the effective theory. These
constraints are not necessarily satisfied when one
defines an effective Hamiltonian by rotating its
matrix in some basis. In the matrix, many
operators may contribute to one and the same
matrix element and a rotation of a matrix does not
have to be equivalent to a rotation of particle
operators. In the effective particle theory, the
matrix rotations one considers result only from
rotating particle operators. This simplification
allows one to encode many matrix elements in a
small number of operators. In practice, the
simplification begins to manifest itself when one
considers states with more particles than 2 or 3.
For example, a two-body effective potential term
involves a product of only four fields and a
function, say $v$, of their four arguments, such
as in Secs. II E and X C in~\cite{Wilsonetal}. In
contrast, a matrix element of the same term
between states of many particles involves many
copies of the function $v$ with different
arguments. In addition, since the field operators
involve both creation and annihilation operators,
even an operator built from just four field
operators requires a large space of states to
study its matrix elements and uncover its actual
operator structure as a result. 

Derivation of the perturbative formulae for
relativistic interactions of effective particles
starts from the non-perturbative RGPEP equation
that has been mentioned before in~\cite{HfRGPEP}
on the basis of~\cite{npRGPEP}. The starting RGPEP
equation resembles Wegner's \cite{Wegner1} (for
reviews, see \cite{Wegner2,Kehrein}) in its double
commutator structure. Somewhat different
perturbative expansions could be obtained in the
RGPEP starting from the non-perturbative equations
that involve multiple commutators. Such equations
are described in Appendix~\ref{MCRGPEP}.

The relativistic nature of effective interactions
is achieved using RGPEP by respecting the 7
kinematical symmetries of the FF of Hamiltonian
dynamics~\cite{DiracFF} and cluster
properties~\cite{Weinberg} in quantum field
theory. The starting non-perturbative equation of
RGPEP is designed to preserve these features. They
are also preserved in the perturbative expansion.
The running cutoff parameter of RGPEP limits only
changes of the invariant mass of interacting
particles. These features are required, for
example, in application to QCD where one desires
to simultaneously explain the constituent quark
model classification of hadrons~\cite{PDG} and
their quite different picture in the parton
model~\cite{partonmodel}. 

Section \ref{NPRGPEP} provides a brief account of
the RGPEP and Section \ref{PF} describes the
perturbative solution up to the 4th order. Section
\ref{C} concludes the paper. Appendix
\ref{MCRGPEP} describes multi-commutator flow
equations and Appendix \ref{AS} provides universal
formulae for coefficients in the perturbative
expansion of effective Hamiltonian interaction
terms.

%%%%%%%%%%%%%%%%%%%%%%%%%%%%%%%%%%%%%%%%%%%%%%%%%
\section{ Non-perturbative RGPEP~\cite{npRGPEP} }
\label{NPRGPEP}
%%%%%%%%%%%%%%%%%%%%%%%%%%%%%%%%%%%%%%%%%%%%%%%%%

Consider a local field theory in which fields on a
given hyper-plane in the Minkowski space-time are
expanded into their Fourier components. The
components have interpretation of creation or
annihilation operators of the field quanta. These
quanta are called the bare, or initial particles,
and the operators that create or annihilate them
are called bare particle operators. The bare
particle operators are generically denoted by
$q_0$. A one-bare-particle state is obtained by
acting with a creation operator $q_0$ on the
vacuum state $|0\rangle$. The bare particle 
operators are used to build a basis in the
Fock-space by acting with products of creation 
operators $q_0$ on $|0\rangle$.

Hamiltonian densities of canonical quantum field
theories are built from products of fields and
their derivatives. A Hamiltonian obtained by
integrating such a density over a space-time
hyperplane is a combination of products of
operators $q_0$ with coefficients $c_0$. If a
product contains $n$ particle operators, the
coefficient has $n$ arguments. Each of the
arguments represents a complete set of quantum
numbers carried by a corresponding particle. One
sums or integrates over these arguments in 
the Hamiltonian terms. 

RGPEP introduces effective particle operators 
\beq
\label{qs}
q_s \es \cU_s \, q_0 \, \cU_s^\dagger \, ,
\eeq
labeled by the parameter $s$ which plays the role
of a renormalization group parameter. This means
that $s$ labels a family of Hamiltonians that all
correspond to one and the same theory but are
expressed in terms of differently defined degrees
of freedom. All kinematical quantum numbers of $q$ 
are the same on both sides of Eq.~(\ref{qs}), i.e., 
irrespective of the value of $s$. 

The parameter $s$ has dimension of length and
ranges from 0 to any finite number of choice,
including arbitrarily large numbers. The larger
$s$ the harder calculation of the corresponding
Hamiltonian. $s \rightarrow \infty$ corresponds to
solving for the spectrum of a theory. This is
exemplified in the elementary model discussed
below. In complex theories, where there is little
hope for obtaining analytic solutions and one has
to use numerical methods, the parameter $s$ may
hopefully be kept, on the one hand, sufficiently
small to maintain an analytically controlled
connection with the quantum field theory one
starts from, and, on the other hand, sufficiently
large to enable numerical calculations using
computers.
 
Physically, $s$ has the interpretation of the
characteristic size of the effective particles.
The interpretation follows from the fact that
effective interactions contain form factors that
limit how far off energy shell the interactions
can extend.\footnote{The energy scale is
introduced using the spectrum of a free part of
the Hamiltonian.} The width of the form factors 
is determined by $1/s$. Consequently, $s=0$
corresponds to the concept of point-like, bare 
particles in the case of a local quantum field 
theory\footnote{When the RGPEP is applied to a 
non-local effective theory, the starting value 
of $s=0$ corresponds to the initial size of 
non-locality.}. The effective Hamiltonian that
corresponds to scale $s$ is band-diagonal on the
energy scale (more precisely, on the scale of
invariant masses of particles involved in an
interaction) and the band width is $\sim 1/s$. The
principle of using the band-diagonal structure for
the purpose of renormalization is formulated in
\cite{GlazekWilson1}.

Let the initial Hamiltonian including
counterterms\footnote{The counterterms have to be
derived and the derivation is a part of the RGPEP
but the initial Hamiltonian is meant to contain
the required counterterms, i.e., the ones to be
found using the RGPEP.} be $\cH_0(q_0)$. For
dimensional and notational reasons, it is
convenient to use the parameter $t=s^4$ and label
Hamiltonians and other operators with $t$ rather
than $s$ itself. The RGPEP demands that 
\beq
\cH_t(q_t) \es \cH_0(q_0) \, .
\eeq
This means that one changes particle operators
from $q_0$ to $q_t$ and at the same time the
coefficients $c_0$ are changed to $c_t$ so that
the Hamiltonian operator stays unchanged.
Consequently, the wave functions of its
eigenstates, in a Fock-space basis built using
$q_t$, depend on $t$. This design serves the
purpose of simultaneously explaining the canonical
picture of Hamiltonian eigenstates as built from
bare, point-like particles, and the constituent
picture in terms of extended, effective particles.
Such setup is desired for solving QCD and
comparing solutions with experimental data.

%%%%%%%%%%%%%%%%%%%%%%%%
\subsection{ Generator }
%%%%%%%%%%%%%%%%%%%%%%%%

By differentiating both sides of
\beq
\cH_t(a_0) \es \cU^\dagger_t \, \cH_0(a_0) \, \cU_t \, ,
\eeq
with respect to $t$, which is denoted by a prime, 
one obtains
\beq 
\label{ht1}
\cH'_t(a_0) \es
[ - \cU_t^\dagger \cU'_t , \cH_t(a_0) ] \, .
\eeq 
The product 
\beq
\label{cG}
\cG_t \es - \cU_t^\dagger \cU'_t 
\eeq
is called a generator. The RGPEP generator is 
defined by the formula
\beq
\label{cGdef}
\cG_t \es [ \cH_f, \cH_{Pt} ] \, ,
\eeq
whose right-hand side will be explained below.
The definition allows for adding arbitrary 
multiples of $\cH_t(a_0)$ or other operators 
that commute with $\cH_t(a_0)$ but such additions
to $\cG_t$ are ignored as immaterial here.

Since $\cG_t$ is defined as a commutator of two
formally hermitian operators, it is formally
anti-hermitian, and $\cU_t$ is formally unitary.
The word ``formal'' is used here because precise
definitions require regularization (see the quoted
literature).

In what follows below, if an operator is expressed 
in terms of the basis operators $q_0$, they are no
longer indicated as its arguments. However, one 
should remember that the effective Hamiltonian is
obtained by making the substitution, $H =
\cH_t(q_t) = \cH_t(q_0 \rightarrow q_t)$. In the
notation for operators $q_0$ themselves, the
subscript $0$ is also omitted, so that $q \equiv
q_0$.

%%%%%%%%%%%%%%%%%%%%%%%%%%%%%%%%%%%%%%%%%%%
\subsection{ $\cH_f$ in Eq. (\ref{cGdef}) }
%%%%%%%%%%%%%%%%%%%%%%%%%%%%%%%%%%%%%%%%%%%

The Hamiltonian $\cH_f$, called the free Hamiltonian, 
is the part of a full Hamiltonian at $t = s^4 = 0$
that has the form 
\beq
\label{cHf} 
\cH_f \es
\sum_i \, p_i^- \, q^\dagger_i q_i 
\eeq 
and does not depend on the interactions. More 
precisely, $\cH_f$ is defined as the part of 
$\cH$ that one is left with when the coupling 
constants in the initial theory are set to 0 
and only the terms that are bilinear in fields 
are kept. The sum extends over all particle 
species and their quantum numbers so that the 
sum also includes integration over particle
momenta $p$. The minus component of a 
particle momentum, $p_i^- = p_i^0 - p_i^3$, 
in Eq. (\ref{cHf}), 
\beq
p^-_i \es { p_i^{\perp \, 2} + m_i^2 \over p_i^+} \, ,
\eeq
is the FF energy that corresponds to a free particle 
with the mass $m_i$ and kinematical momentum components 
$p_i^+$ and $p_i^\perp$. Typically, different species 
of particles have different masses.

%%%%%%%%%%%%%%%%%%%%%%%%%%%%%%%%%%%%%%%%%%%%%%
\subsection{ $\cH_{Pt}$ in Eq. (\ref{cGdef}) }
%%%%%%%%%%%%%%%%%%%%%%%%%%%%%%%%%%%%%%%%%%%%%%

The operator $\cH_{Pt}$ is uniquely defined once
the Hamiltonian $\cH_t$ is specified. If the 
Hamiltonian $\cH_t$ is of the form 
\beq
\label{Hstructure} 
{\cal H}_t \es
\sum_{n=2}^\infty \, 
\sum_{i_1, i_2, ..., i_n} \, c_t(i_1,...,i_n) \, \, q^\dagger_{i_1}
\cdot \cdot \cdot q_{i_n} \, ,
\eeq 
where the coefficients $c_t(i_1,...,i_n)$ are to 
be found using RGPEP, the operator $\cH_{Pt}$ is 
defined by
\beq
\label{HPstructure} 
{\cal H}_t \es
\sum_{n=2}^\infty \, 
\sum_{i_1, i_2, ..., i_n} \, c_t(i_1,...,i_n) \, \,\left( {1 \over
2}\sum_{k=1}^n p_{i_k}^+ \right)^2 \, \, q^\dagger_{i_1}
\cdot \cdot \cdot q_{i_n} \, .
\eeq 
Thus, ${\cal H}_{Pt}$ differs from ${\cal
H}_t$ by multiplication of its terms by the
square of $+$-component of total momentum 
carried by the particles in a term. This 
kinematical momentum is specified by the 
operator content of a term irrespective
of the value of RGPEP parameter $t = s^4$.

The definition of $\cH_{Pt}$ secures that 
both sides of Eq. (\ref{ht1}), which now 
reads
\beq 
\label{tnpRGPEP}
{d \over dt} \, \cH_t \es
\left[ [ \cH_f, \cH_{Pt} ], 
\cH_t \right] \, ,
\eeq 
behave in the same way with respect to operations
of kinematical LF symmetries. This implies that
the effective particle size parameter $s$ is
invariant with respect to the FF kinematical
subgroup of the Poincar\'e group. 

%%%%%%%%%%%%%%%%%%%%%%%%%%%%%%%%%%%%%%%%%%%
\subsection{ Construction of counterterms }
%%%%%%%%%%%%%%%%%%%%%%%%%%%%%%%%%%%%%%%%%%%

The RGPEP Eq. (\ref{tnpRGPEP}) predicts the
coefficients $c_t$ of products of $q_t$ in
$\cH_t(q_t)$ provided the initial condition for
$\cH_0(q_0)$ is available. However, when the
initial Hamiltonian involves divergences, such as
the ones due to a dynamical coupling of infinitely
many degrees of freedom over an infinite range of
scales, it cannot be considered a good initial
condition. One can cut the infinities off by
regularization, limiting the space of states
and/or limiting the range of scales involved in
interaction. With the cutoffs, the solutions lead
to effective Hamiltonians that involve huge
numbers (or zeros) instead of meaningful terms.
The huge numbers result from the ratios of cutoffs
to physical parameters. Zeros result from inverses
of the huge ratios. 

The situation is understood here as the one in
which the initial Hamiltonian may reflect a
physically interesting structure that is relevant
when one imposes so small cutoffs on the dynamics
that the divergences are replaced by relatively 
small terms and these terms do not obscure the 
initial structure beyond recognition. The question 
is then how a good effective Hamiltonian should 
depend on the cutoffs so that its predictions do 
not. If the predictions did depend on the cutoffs, 
one could not treat the cutoff theory as equivalent 
to some fundamental one. By the word ``fundamental'' 
it is meant here that the range of validity of a 
``fundamental'' theory is expected to be vastly 
greater than the range limited by adjustable 
cutoffs.

Following~\cite{Wilson1,Wilson2}
and~\cite{GlazekWilson1}, the RGPEP involves
determination of additional terms, called
counterterms, that need to be included in the
initial Hamiltonian in order to render the
effective theory which may correspond to the
Hamiltonian initially suggested as valid. The
construction of counterterms follows the rules of
SRG procedure~\cite{GlazekWilson1, RGPEP,npRGPEP}.
There is no explicit condition of widening of the
band toward high energies (large invariant
masses), but the narrowness of the Hamiltonian
(see next section) enables one to search for
counterterms on the basis of a condition that all
matrix elements of a Hamiltonian corresponding to
some finite size $s$ of effective particles
between states of finite invariant masses, are
independent of the initial cutoffs. This condition
provides as many equations to solve as there are
matrix elements to consider. The number of
equations may be large enough to specify the
structure of counterterms up to a limited set of
unknown finite numbers or functions. These numbers
or functions, called finite parts of counterterms,
can be constrained by symmetries~\cite{PerryWilsonCC} 
and phenomenology.

In general, the only method for solving
Eq.~(\ref{tnpRGPEP}) is to use successive
approximations. One hopes that a candidate number
$n$ for a solution, $\cH_t^{(n)}$, can be inserted
on the right-hand side to render a new candidate
$\cH_t^{(n+1)}$ on the left-hand side, and
$\cH_t^{(n+1)}$ can be subsequently inserted on
the right-hand side in place of $\cH_t^{(n)}$ to
render $\cH_t^{(n+2)}$ on the left-hand side, and
so on. More precisely, the hope is that a
well-defined solution is approximated with
increasing accuracy when $n$ increases. Of course,
one has to study theories case-by-case in order to
establish if the RGPEP sequence
converges~\cite{WilsonFisher}. In principle, one
may obtain not only solutions for $\cH_t$ with
limit cycles~\cite{lc1}, instead of just fixed
points, but also the solutions with chaotic
behavior as functions of $t$~\cite{lc2,lcet}. 

The conceptual and computational difficulty is
that in each step of constructing successive
approximations one may have to modify the
counterterms and the modification is not dictated
by the iteration procedure alone. Namely, one has
to inspect dependence of the small-invariant-mass
matrix elements of $\cH_t^{(n)}$ at large $t$ on
regularization present in $\cH_t^{(n)}$ at $t=0$
and attempt to counter this dependence by
modifying the $\cH_t^{(n)}$ at $t=0$. The
modifications may amount to redefinitions of a
finite number of coupling constants or functions
of particle quantum numbers. However, when one
confronts the issue of relativistic description of
confinement, the question of convergence is not
answered in any form yet and the mechanism by
which effective Hamiltonians develop confining
forces at finite $t$ is strictly speaking not
known.

The attractive feature of the RGPEP equation is
that it allows for splitting of the integration
process into many small steps, accompanied with
rescaling of invariant masses so that one always
operates with dimensionless quantities $ s \cM $,
where $\cM$ is a free invariant mass of the
interacting effective particles and $s$ is their
size parameter. Each small step can be executed
using variety of well-known mathematical methods
as in the case of the original Wilsonian procedure
\cite{Wilson1,Wilson2}. Thus, the RGPEP provides a
new tool for studying universality in relativistic
quantum field theories (see Appendix A
in~\cite{npRGPEP}). This feature is of interest
because a well-defined effective QCD Hamiltonian
may be sought using the RGPEP irrespective of many
details in the initial Hamiltonian. Namely, it
should be sufficient to start with any Hamiltonian
in a suitable universality class.

Nevertheless, in order to begin the process of
constructing solutions of the RGPEP equations as
outlined above, one has to suggest the terms that
should be included in $\cH_t^{(1)}$ and can be
expected to have significant coefficients in
$\cH_t^{(n)}$ with large $n$. Initially, the only
available tool for gathering such information
about important terms is perturbation theory. This
article illustrates a general algorithm for
generating perturbative formulae for $\cH_t$. The
illustration includes a set of perturbative
formula up to 4th order, derived in Section
\ref{PF}. 

%%%%%%%%%%%%%%%%%%%%%%%%%
\subsection{ Narrowness }
\label{Narrow}
%%%%%%%%%%%%%%%%%%%%%%%%%

Solving Eq. (\ref{tnpRGPEP}) by successive
approximations involves restrictions on how many
terms are kept in the Hamiltonian. Each of these
terms is written as a product of quantum fields. 
A priori, a FF $\cH_t$ contains infinitely many
terms~\cite{Wilsonetal}. A good approximate
operator solution must be little sensitive to
the terms that are missing. Assuming that one can
identify a set of dominant operators, the
mechanism of narrowing the Hamiltonians through
Eq. (\ref{tnpRGPEP}) can be seen using an equation
that results from projecting Eq. (\ref{tnpRGPEP})
on a subspace of the Fock space. 

Following Appendix C in Ref.~\cite{npRGPEP}, 
one can introduce a projector on a subspace 
of physical interest in the Fock space. Both
the subspace and projector are denoted by
$R$. The projection involves fixing total 
kinematical momentum of states and a cutoff 
on their total invariant mass so that the 
concept of a trace becomes well-defined. Let 
\beq 
\cH \es R \, \cH_t R \, .
\eeq
The correspondingly projected RGPEP equation 
reads
\beq 
\label{narrowR2}
\cH' \es
\left[ [\cH_f, \cH_P], \cH \right] \, .
\eeq 
The condition that the trace of $\cH^2$ does
not depend on $t$, implies for matrix elements
$\cH_{mn} = \langle m | \cH | n \rangle$ in the 
basis built from eigenstates $|m\rangle$ of 
$\cH_f$ with eigenvalues $P^-_{fm}$, that 
\beq
\label{start5x}
\left( \sum_{mn} |\cH_{Imn}|^2 \right)'
\es
- \sum_{km} (P^-_{fk} - P^-_{fm})^2 |\cH_{Ikm}|^2
2P_{mk}^{+2} \le 0 \, ,
\eeq
where $P^+_{mk}$ is the total $+$-momentum 
of the particles that are involved in the
interaction. In terms of the invariant masses,
\beq
\label{start6x}
\left( \sum_{mn} |\cH_{Imn}|^2 \right)'
\es
- 2 \sum_{km} (\cM^2_{km} - \cM^2_{mk})^2
|\cH_{Ikm}|^2 \le 0 \, ,
\eeq
where $\cM_{km}$ denotes an invariant mass of
the particles in state labeled with $k$ that 
are transformed through the interaction $\cH_I$  
to particles in the state labeled by $m$. 
Change of order of subscripts results in an
invariant mass of particles in a different state.
Spectators do not contribute to the invariant 
masses. The calculations are explicitly invariant 
under 7 kinematical transformations of the FF 
of Hamiltonian dynamics.

Eq. (\ref{start6x}) means that the sum of moduli 
squared of all matrix elements of the interaction 
Hamiltonian decreases as $t$ increases until all 
off-diagonal matrix elements of the interaction 
Hamiltonian between states with different free 
invariant masses vanish. Section \ref{PF} shows 
this feature in a perturbative expansion for
$\cH_t$ in powers of $\cH_{It}$. 

%%%%%%%%%%%%%%%%%%%%%%%%%%%%%%%%%%%%%
\subsection{ Transformation $\cU_t$ }
%%%%%%%%%%%%%%%%%%%%%%%%%%%%%%%%%%%%%

Transformation $\cU_t$ is a solution to
\beq
\cU_t'
\es 
- \cU_t \,\, [ \cH_f, \cH_{Pt} ] \, .
\eeq
The initial condition is $\cU_0 = 1$. 
Successive approximations generate 
$\cU_t$ from a solution for $\cH_t$. 
The general solution is 
\beq
\label{Usolution}
\cU_t 
\es 
T \exp{ \left( - \int_0^t d\tau \, [ \cH_f, \cH_{P\tau} ] \right) } \, ,
\eeq
where $T$ orders operators from left to right 
in the order from a smallest to a largest of 
their arguments $t$. Perturbative expansion of 
Eq. (\ref{Usolution}) provides a perturbative 
solution for $\cU_t$. Knowing $\cU_t$, one can
calculate operators that create or annihilate 
effective particles of size $s$ using Eq. 
(\ref{qs}) with $t=s^4$.

%%%%%%%%%%%%%%%%%%%%%%%%%%%%%%%%%
\section{ Perturbative formulae }
\label{PF}
%%%%%%%%%%%%%%%%%%%%%%%%%%%%%%%%%

Solution of Eq. (\ref{tnpRGPEP}) for coefficients
of products of effective particle operators in
$\cH_t$, is described in this Section using the
notation that proved useful before\footnote{E.g., 
see~\cite{GlazekScalar}, Sec. II.B, or~\cite{gluons}, 
Sec. III.C.}, in which Eq. (\ref{tnpRGPEP}) takes 
the form
\beq 
\label{CtnpRGPEP}
\cH_{t \, ab}' \es
- ab^2 \, \cH_{It \, ab} 
+ \sum_x (p_{ax} \, ax + p_{bx} \, bx) \, 
\cH_{It \, ax} \, \cH_{It \, xb} \, .
\eeq 
The letters $a$, $b$, and $x$, denote configurations 
of particles. A configuration is a collection of 
quantum numbers that label particle operators.

\begin{figure}
\begin{center}
\includegraphics[scale=.3]{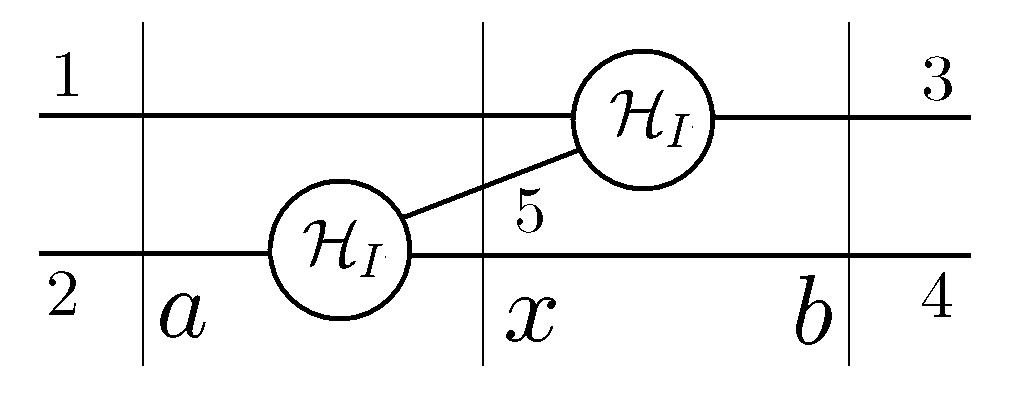}
\end{center}
\caption{ Example of the RGPEP operator calculus, 
see Eq. (\ref{CtnpRGPEP}) and its description. }
\label{Fig1}
\end{figure}

An example illustrating the concept of
configuration is shown in Fig.~\ref{Fig1}. In
Fig.~\ref{Fig1}, the configuration $a$ includes
particle operator labels contained in the sets
denoted by 1 and 2, the configuration $x$ includes
particle operator labels contained in the sets
denoted by 1, 5 and 4, and the configuration $b$
includes particle operator labels contained in the
sets denoted by 3 and 4. A subscript such as $ax$
refers to a coefficient of the product of particle
operators in a Hamiltonian term $\cH_I$ that
changes a configuration $x$ to a configuration
$a$, etc. In Fig. \ref{Fig1}, $\cH_{It \, ax}$
would correspond to a term in $\cH_{It}$ that
involves a product of creation operators labeled
with quantum numbers in set 2 and a product of
annihilation operators with labels contained in
sets 5 and 4. $\cH_{It \, xb}$ would correspond to
a term that involves a product of creation
operators labeled with quantum numbers in sets 1
and 5 and a product of annihilation operators with
labels contained in set 3. Symbol $ax$ as a term
in the equation denotes a difference of invariant
masses squared, $ax = \cM^2_{ax} - \cM^2_{xa}$.
$\cM_{ax}$ denotes an invariant mass corresponding
to only those particle operator labels in a
configuration $a$ that emerge as a result of
interaction from the configuration $x$. Thus, $ax
= - xa$. For example, in Fig.~\ref{Fig1},
$\cM_{ax}$ denotes the invariant mass
corresponding to the particle operator labels in
set 2 and $\cM_{xa}$ denotes the invariant mass
corresponding to the particle operator labels in
sets 5 and 4. Spectators do not contribute to
these invariant masses. Thus, in Fig.~\ref{Fig1},
$\cM_{ax}$ and $\cM_{xa}$ do not include particle
operator labels from set 1, while $\cM_{xb}$ and
$\cM_{bx}$ do not include particle operator labels
from set 4. The invariant masses are evaluated
using kinematical momenta and eigenvalues of
$\cH_f$. A symbol such as $p_{ax}$ denotes the sum
of $+$ components of all momenta that appear in
the labels of creation operators (or of all
annihilation operators, which is the same due to
the momentum conservation) in a product in the
Hamiltonian term that changes the configuration
$x$ to $a$, etc. In Fig.~\ref{Fig1}, $p_{ax}$
denotes the sum of $+$ components of momenta
contained in set $2$, which is the same as the sum
of $+$ components of momenta contained in sets 5
and 4 together, while $p_{xb}$ denotes the sum of
$+$ components of momenta contained in sets 1 and
5, which is the same as the sum of $+$ components
of momenta contained in set~3.

Key features of Eq. (\ref{CtnpRGPEP}) are
following. Since Eq. (\ref{CtnpRGPEP}) only refers
to quantum numbers that label particle operators,
the Hamiltonian operators that solve Eq.
(\ref{CtnpRGPEP}) act in the entire Fock space.
This is how the RGPEP procedure avoids the
limitation to an a priori limited set of states
that one is forced to work with using equations
for Hamiltonian matrices. There are no
disconnected terms in $\cH_t$, since the second
term in Eq. (\ref{CtnpRGPEP}) results from a
commutator. The commutator also implies that one
keeps only the terms that result from commuting at
least one annihilation operator with one creation operator
in the product $\cH_{It \, ax} \, \cH_{It \, xb}$.
Also, $\cH_t$ does not explicitly depend on the
eigenvalues of a full Hamiltonian. Such dependence
is a serious limitation of all procedures that are
in principle based on Gaussian elimination, or, in
the path integral formulation, on integrating out
the states outside a cut off in the space of
states~\cite{Wilson2}. Namely, the larger an
eigenvalue the worse accuracy of the procedure for
integrating out the components above a cutoff. 

The problem of integrating out states above a
cutoff is circumvented in the RGPEP at the level
of constructing counterterms and evaluating
effective Hamiltonians. A finite remnant of this
problem remains in an eigenvalue equation for
$\cH_t(q_t)$, when one limits the space of states
of effective particles that are included in a
non-perturbative diagonalization using computers
and when one fits finite parts of counterterms by
comparison of theoretical results with experiment.
This residual issue must be dealt with in the
diagonalization procedure and typically should not
cause problems because interactions do not grow
with energy fast enough to generate any
divergences. If they did cause divergences, a
separate search for fixed points or other in
principle possible behavior would have to be
arranged (see Appendix B in~\cite{GlazekWilson1}).

%%%%%%%%%%%%%%%%%%%%%%%%%%%
\subsection{ Form factors }
\label{ff}
%%%%%%%%%%%%%%%%%%%%%%%%%%%

Eq. (\ref{CtnpRGPEP}) suggests that one writes
\beq
\cH_{t \, ab} \es e^{ - t\, ab^2 } \,\, \cG_{t \, ab} \, ,
\eeq
where 
\beq
f_{t \, ab} \es e^{ - t\, ab^2 } 
\eeq
is called a form factor. The width of the form factor 
as a function of the invariant mass change is $1/s$. 
Therefore, the parameter $s$ has the interpretation of 
a size of effective particles. It determines how far
off shell an interaction can reach in terms of the 
invariant mass. The form factor secures narrowness
of the effective Hamiltonian order-by-order in
perturbation theory.

Eq. (\ref{CtnpRGPEP}) implies that $\cG_{t \, ab}$ 
satisfies the equation
\beq 
\label{cgprime}
\cG_{t \, ab}' \es
\sum_x A_{t \, axb} \, \cG_{t \, ax} \, \cG_{t \, xb} \, , 
\eeq
where
\beq
\label{Ataxb}
A_{t \, axb} 
\es
(p_{\,ax} \, ax + p_{\,bx} \, bx) \, e^{- t( ax^2 + xb^2 - ab^2)} \, .
\eeq
Eq. (\ref{cgprime}) is solved below order-by-order 
in a perturbative expansion using powers of a single
coupling constant $g$.

%%%%%%%%%%%%%%%%%%%%%%%%
\subsection{ Expansion }
%%%%%%%%%%%%%%%%%%%%%%%%

When one expands 
\beq
\cG_t \es \cH_f + g \, \cG_{t1} + g^2 \, \cG_{t2} + g^3 \, \cG_{t3} + ... \, ,   
\eeq
the perturbative formulae for calculating
$\cG_{tn}$ is obtained in the form 
\beq 
\label{cgprime1}
\cG_{t \, n \, ab}' \es
\sum_{k=1}^{n-1} \sum_x A_{t \, axb} \, \cG_{t \, k \, ax} \, \cG_{t \, n-k  \, xb} \, .
\eeq
It yields order-by-order the following differential equations 
\beq 
\label{1st}
\cG'_{t1 \, ab} \es 0 \, , \\
\label{2nd}
\cG'_{t2 \, ab} \es \sum_x A_{t \, axb} \, 
\cG_{t1 \, ax} \, \cG_{t1 \, xb} \, , \\
\label{3rd}
\cG'_{t3 \, ab} \es \sum_x A_{t \, axb} \, 
\left( \cG_{t1 \, ax} \, \cG_{t2 \, xb} 
+ \cG_{t2 \, ax} \, \cG_{t1 \, xb} \right) \, , \\ 
\label{4th}
\cG'_{t4 \, ab} \es \sum_x A_{t \, axb} \, 
\left( \cG_{t1 \, ax} \, \cG_{t3 \, xb} 
     + \cG_{t2 \, ax} \, \cG_{t2 \, xb}     
     + \cG_{t3 \, ax} \, \cG_{t1 \, xb} \right) \, .
\eeq
The four terms indicate a generic pattern. This 
pattern is different from the patterns described
in perturbative expansions studied using matrix 
models in Refs.~\cite{AlteredWegner} and 
\cite{AccuracyEstimate}. The main reasons for the 
patterns to differ are the constancy of $\cH_f$, 
which is of order 1, and the definition of $\cH_P$ 
in Eq. (\ref{tnpRGPEP}), which leads to the factors 
of $+$-momentum of active particles in Eq. (\ref{Ataxb}). 

The same factors of $p^+$ appear in the procedure
studied in Ref.~\cite{AccuracyEstimate}, but there
they are accompanied by appearance of the
derivative of $\cG'$ on the right-hand side of the
RGPEP equation for $\cH'$, in the generator. The
derivatives in the generator cause additional
difficulties in solving the RGPEP equation
non-perturbatively. Here, the origin of factors of
$p^+$ is in $\cH_{Pt}$. The absence of derivatives
in the generator remedies the additional difficulties 
encountered in Ref.~\cite{AccuracyEstimate}.

General formulae for solutions of Eq.
(\ref{cgprime1}) are listed below order-by-order
for all terms up to the 4th order, i.e., up to 2
or 3 loops in theories of the type $\phi^3$ or
$\phi^4$, respectively, where the power of $\phi$
indicates how many field operators can be present
in a term. This covers in principle all cases of
interest in physics (except for gravity, because
of the FF limitation to the Minkowski metric). For
example, the expansion to 4th order is in
principle sufficient to study interactions of
constituent quarks in hadrons including the
leading effects of self-interaction, gluon
exchange, and running coupling constant. Such
studies should clarify if the 4th order RGPEP is 
capable of generating information about the 
essentially non-canonical effective QCD interaction 
terms discussed in Ref.~\cite{Wilsonetal}, or
one must explore orders higher than 4 to see the
new terms. Terms of higher orders can be generated
as required according to the pattern visible in
what follows. 

%%%%%%%%%%%%%%%%%%%%%%%%%%%%%%%%%%%%%%%%
\subsection{ Solutions, order-by-order }
\label{S}
%%%%%%%%%%%%%%%%%%%%%%%%%%%%%%%%%%%%%%%%
\noindent
Appendix \ref{AS} describes details of solving 
Eqs.~(\ref{1st}) to (\ref{4th}). This section
lists the results, introducing the elements 
of notation developed in Appendix \ref{AS}.

%%%%%%%%%%%%%%%%%%%%%%%%%%%%%%%%%%%
\subsubsection{ 1st-order solution }
%%%%%%%%%%%%%%%%%%%%%%%%%%%%%%%%%%%
\noindent
The first-order solution does not depend on $t$, 
\beq 
\label{s1st}
\cG_{t1 \, ab} \es \cG_{01 \, ab}  \, .
\eeq
The subscript 0 refers to $t=0$. Operators
with subscript 0 are contained in the initial
Hamiltonian. They provide the initial conditions.

The first-order initial condition includes a bare
coupling constant, $g$, as a coefficient in front
of a definite canonical operator. Counterterms of
higher orders typically include the same operator.
The net result of including counterterms is a 
change of value of $g$. The magnitude of change 
depends on the interaction terms included in and 
regularization method adopted for the initial 
canonical Hamiltonian.

%%%%%%%%%%%%%%%%%%%%%%%%%%%%%%%%%%%%
\subsubsection{ 2nd-order solution }
%%%%%%%%%%%%%%%%%%%%%%%%%%%%%%%%%%%%
\noindent
The second-order solution reads
\beq 
\label{s2nd}
\cG_{t2 \, ab}  \es \cG_{02 \, ab} + \sum_x
B^{(123,0)}_{t \, axb} \, \cG_{02 \, axb} \, .
\eeq
The initial condition $\cG_{02 \, ab}$ typically
includes instantaneous interactions, such as a FF
counterpart of the Coulomb interaction term in the
standard form of dynamics, or instantaneous
fermion terms that are unique to the FF of
dynamics, due to constraints. The RGPEP may
provide the higher-order terms that have similar
structure but whose coefficients are not
necessarily constrained according to the simple
Heisenberg equations of motion that one might
expect from the analogy with classical field
equations \cite{DiracDeadWood}. For example,
quantum interactions may lead to anomalies. The
initial condition $\cG_{02 \, ab}$ also includes
self-interaction counterterms. 

The operator structure $\cG_{02 \, axb}$ is
defined in Eq. (\ref{02axb}). It results from
action of first-order terms twice. The coefficient
$B^{(123,0)}$ is defined in Eq. (\ref{(123,0)}).
The superscript convention is explained in
Appendix \ref{notationsummary}. It is well-known
that the second-order terms reproduce standard
second-order results for observables to the extent
they are available in all theories relevant to
physics. 

%%%%%%%%%%%%%%%%%%%%%%%%%%%%%%%%%%%%
\subsubsection{ 3rd-order solution }
%%%%%%%%%%%%%%%%%%%%%%%%%%%%%%%%%%%%
\noindent
The third-order solution reads
\beq
\label{s3rd}
\cG_{t3 \, ab} 
\es 
\cG_{03 \, ab}
+
\sum_x  B^{(123,0)}_{t \, axb} \, \cG_{03 \, axb} 
\np
\sum_{xy} \, \left[ B^{(124,(234,0))}_{t \, axyb} +
                    B^{(134,(123,0))}_{t \, axyb} \right] \, \cG_{03 \, axyb}
\, . 
\eeq
The term $\cG_{03 \, ab}$ is an initial condition 
for an operator that does not appear in a canonical 
Hamiltonian. Namely, it is a third-order coupling 
constant counterterm. Operators $\cG_{03 \, axb}$ 
involve a product of two operators from the initial 
Hamiltonian: one of first order and another one 
of second order, see Eq. (\ref{03axb}). The last 
operator, $\cG_{03 \, axyb}$, is a product of three
first-order operators, see Eq. (\ref{03axyb}).
The coefficients $B^{(124,(234,0))}_{t \, axyb}$
and $B^{(134,(123,0))}_{t \, axyb}$ are defined 
in Eqs. (\ref{(124,(234,0))}) and (\ref{(134,(123,0))})
in Appendix \ref{notationsummary}. 

The third-order solution of Eq. (\ref{s3rd})
slightly differs from the third-order formula
previously studied using RGPEP in the cases of
scalar theories in 6
dimensions~\cite{GlazekScalar} and in QCD in 4
dimensions~\cite{gluons}. Although the previous
experience suggests that Eq. (\ref{s3rd}) leads to
the same result concerning asymptotic freedom, the
finite invariant-mass details of Eq. (\ref{s3rd}) 
are of special interest since they contribute to 
the fourth-order terms. The fourth-order terms 
required prohibitively complex equations in the 
previous applications of RGPEP. A relatively 
compact fourth-order solution is now made available 
in the next section.

%%%%%%%%%%%%%%%%%%%%%%%%%%%%%%%%%%%%
\subsubsection{ 4th-order solution }
\label{4s}
%%%%%%%%%%%%%%%%%%%%%%%%%%%%%%%%%%%%

The fourth-order solution reads
\beq
\label{G4}
\cG_{t4 \, ab} 
\es 
\cG_{04 \, ab} 
+
\sum_x 
B^{(123,0)}_{t \, axb} \, \cG_{04 \, axb}
\np
\sum_{xy} 
\left[ B^{(124,(234,0))}_{t \, axyb} +
B^{(134,(123,0))}_{t \, axyb} \right] \,
\cG_{04 \, axyb}
\np
\sum_{xyz} 
\left[
B^{(125,(235,(345,0)))}_{t \, axyzb} 
+
B^{(125,(245,(234,0)))}_{t \, axyzb} 
+
B^{(145,(124,(234,0)))}_{t \, axyzb} 
\right.
\np
\left.
B^{(145,(134,(123,0)))}_{t \, axyzb} 
+
B^{(135,(123,0),(345,0))}_{t \, axyzb} 
\right] \, \cG_{04 \, axyzb} \, .
\eeq
The initial condition $\cG_{04 \, ab}$ is the
fourth-order counterterm. It includes the
self-interaction and coupling-constant
counterterms that involve additional particles in
an interaction. It also includes box-diagram
counterterms, where they are needed. The operator
$\cG_{04 \, axb}$ is defined in Eq. (\ref{04axb}),
the operator $\cG_{04 \, axyb}$ is defined in Eq.
(\ref{04axyb}), and the operator $\cG_{04 \,
axyzb}$ is defined in Eq.(\ref{04axyzb}). The new
coefficients in fourth order terms are defined in 
Eqs. (\ref{(125,(235,(345,0)))}) to
(\ref{(135,(123,0),(345,0))}). 

The general fourth-order result opens a door to 
many specific studies. Perhaps one of the most 
instructive ones would be a calculation of 
$\Upsilon$ family masses and decay width, which 
is likely to shed some light on the dynamics of 
gluons in heavy quarkonia, cf.~\cite{HQ}.

%%%%%%%%%%%%%%%%%%%%%%
\section{ Conclusion }
\label{C}
%%%%%%%%%%%%%%%%%%%%%%

The perturbative RGPEP formulae of Section
\ref{PF} allow one to study low-order divergences
and effective interactions in specific theories
with infinitesimal coupling constants and extreme
cutoffs in their canonical Hamiltonians. They also
provide a tool for testing the formal RGPEP claim
of narrowness of effective theories described in
Section \ref{NPRGPEP}. Such tests are essential
because of anomalies, i.e., the new Hamiltonian terms
that do not correspond to classical Lagrangians
and instead result from imposing some regularization
on a formal quantum theory. One potential source
of anomalies that needs to be studied is a lower 
bound on $+$-momenta of particles in a Fourier 
analysis of a quantum field. Strictly speaking, 
such lower bound violates boost invariance of a 
formal theory. 

Regarding systematic checks of existence of
anomalies in the FF of Hamiltonian dynamics, there
is currently no alternative known to the author to
inspecting theories order-by-order using the RGPEP
and finding out term-by-term if and how soon a
specific regularization method in a given theory
may generate anomalies in the weak-coupling
expansion. Moreover, there is currently nothing
known yet for certain about the anomalies
generated in the FF of Hamiltonian dynamics using
RGPEP. 

The subject certainly deserves a study in view of 
the fact that the FF vacuum problem is differently 
formulated from the instant form vacuum problem. 
For example, if one imposes the above mentioned 
lower bound on $+$-momenta of particles in a Fourier 
analysis of quantum fields, the mathematical vacuum 
state, i.e., the state that is annihilated by bare 
annihilation operators, is an exact eigenstate of 
the full Hamiltonian with eigenvalue 0. If this
state can play the role of the true ground state, 
the mechanisms of symmetry breaking and mass 
generation in effective theories may in principle 
turn out not to be associated with the vacuum 
expectation values of normal-ordered products 
of quantum fields but with some new terms in the 
Hamiltonians~\cite{Wilsonetal}. The perturbative 
RGPEP formulae for relativistic interactions of 
effective particles derived in Section \ref{S} 
provide a new tool for studies of such hypotheses. 

Most interestingly, however, the perturbative
studies may help in identifying a finite set of
operators that one can use to approximately solve
the RGPEP Eq. (\ref{CtnpRGPEP})
non-perturbatively, i.e., in terms of the
coefficients in front of the identified operators.
As functions of $t$, these coefficients would be
expected to evolve from a set of constants and
regulating functions at $t=0$ to a set of finite,
non-trivial functions of particle quantum-numbers
at some $t_0 = s_0^4$, where $s_0$ denotes some
size of the effective constituents that are most
suitable as degrees of freedom for specific 
purposes of phenomenology.

\begin{appendix}

%%%%%%%%%%%%%%%%%%%%%%%%%%%%%%%%%
\section{ Multicommutator RGPEP }
\label{MCRGPEP}
%%%%%%%%%%%%%%%%%%%%%%%%%%%%%%%%%

One can obtain narrow relativistic Hamiltonians 
using equations with an odd number of commutators. 
Namely, one can write 
\beq 
\label{MtnpRGPEP}
{d \over dt} \, \cH_t \es
\left[ [ \cH_f ,[...\cH_f,[ \cH_f, \cH_{Pt} ]...]], 
\cH_t \right] \, ,
\eeq 
including $2 n - 1$ operators $\cH_f$ in a sequence 
of commutators with $n > 1$ and $t = s^{4n}$. 
Correspondingly, one can define the operator 
$\cH_{Pt}$ by writing
\beq
{\cal H}_t \es
\sum_{n=2}^\infty \, 
\sum_{i_1, i_2, ..., i_n} \, c_t(i_1,...,i_n) \, \,
\left( {1 \over 2}
\sum_{k=1}^n p_{i_k}^+ \right)^{2n} \, \, q^\dagger_{i_1}
\cdot \cdot \cdot q_{i_n} \, ,
\eeq 
instead of Eq. (\ref{HPstructure}). The resulting 
counterpart of Eq. (\ref{CtnpRGPEP}) reads 
(cf.~\cite{npRGPEP}, footnote 43)
\beq 
\label{MCtnpRGPEP}
\cH_{t \, ab}' \es
- ab^{2n} \cH_{It \, ab} 
+ \sum_x (p_{ax} ax^{2n-1} + p_{bx} bx^{2n-1}) \, 
\cH_{It \, ax} \cH_{It \, xb} .
\eeq 
The narrowness is obtained for the matrices 
of projected Hamiltonians as in Section \ref{Narrow}. 
However, instead of condition (\ref{start6x}), one obtains 
a similar inequality with $(\cM^2_{km} - \cM^2_{mk})^{2n}$ 
in place of $(\cM^2_{km} - \cM^2_{mk})^2$. The 
perturbative expansion proceeds without any
qualitative change, but the form factors in
Section \ref{ff} are of the form
\beq
f_{t \, ab} \es e^{ - t\, ab^{2n} } \, .
\eeq

%%%%%%%%%%%%%%%%%%%%%%%%%%%%%%%%%%%%%%%%%%%%%%%%%
\section{ Calculation of the 4th-order terms }
\label{AS}
%%%%%%%%%%%%%%%%%%%%%%%%%%%%%%%%%%%%%%%%%%%%%%%%%

This Appendix contains a derivation of the
fourth-order term described in Section~\ref{S}.
It also explains the pertinent notation. The calculation
involves all terms of orders lower than 4. Using
solutions given in Eqs.~(\ref{s1st}), (\ref{s2nd}), 
and (\ref{s3rd}), one obtains from Eq.~(\ref{4th})
that 
\beq
\cG'_{t4 \, ab} 
\es 
\sum_x 
A_{t \, axb} \, 
\left( \cG_{01 \, ax} \, \cG_{03 \, xb}
     + \cG_{03 \, ax} \, \cG_{01 \, xb}
     + \cG_{02 \, ax} \, \cG_{02 \, xb} \right)
\np
\sum_{xy} 
A_{t \, axb} \, B^{(123,0)}_{t \, xyb} \, 
\left( \cG_{01 \, ax} \, \cG_{03 \, xyb} 
     + \cG_{02 \, ax} \, \cG_{02 \, xyb} \right)
\np
\sum_{xy} 
A_{t \, ayb} \, B^{(123,0)}_{t \, axy} \, 
\left( \cG_{03 \, axy} \, \cG_{01 \, yb} 
     + \cG_{02 \, axy} \, \cG_{02 \, yb} \right)
\np
\sum_{xyz} 
A_{t \, axb} \, \left[ B^{(124,(234,0))}_{t \, xyzb} + B^{(134,(123,0))}_{t \, xyzb} \right] \, 
\cG_{01 \, ax} \, \cG_{03 \, xyzb} 
\np
\sum_{xyz} 
A_{t \, azb} \, \left[ B^{(124,(234,0))}_{t \, axyz} + B^{(134,(123,0))}_{t \, axyz} \right] \, 
\cG_{03 \, axyz} \, \cG_{01 \, zb}
\np
\sum_{xyz} 
A_{t \, ayb} \, B^{(123,0)}_{t \, axy} \, B^{(123,0)}_{t \, yzb} \, 
\cG_{02 \, axy} \, \cG_{02 \, yzb} 
 \, ,
\eeq
where
\beq
\cG_{02 \, axb} \es \cG_{01 \, ax} \, \cG_{01 \, xb} \, , \\
\cG_{03 \, axb} 
\es  
\cG_{01 \, ax} \, \cG_{02 \, xb} + \cG_{02 \, ax} \, \cG_{01 \, xb} \, , \\
\cG_{03 \, axyb} 
\es  
\cG_{01 \, ax} \, \cG_{01 \, xy} \, \cG_{01 \, yb} \, .
\eeq
Inserting these definitions and grouping coefficients 
in front of the same operators, one obtains the 
derivative of fourth-order terms in the form
\beq
\cG'_{t4 \, ab} 
\es 
\sum_x 
A_{t \, axb} \, 
\left( \cG_{01 \, ax} \, \cG_{03 \, xb}
     + \cG_{03 \, ax} \, \cG_{01 \, xb}
     + \cG_{02 \, ax} \, \cG_{02 \, xb} \right)
\np
\sum_{xy} 
\left[ A_{t \, axb} \, B^{(123,0)}_{t \, xyb} 
     + A_{t \, ayb} \, B^{(123,0)}_{t \, axy} \right]
\nt
\left( \cG_{01 \, ax} \, \cG_{01 \, xy} \, \cG_{02 \, yb} 
     + \cG_{01 \, ax} \, \cG_{02 \, xy} \, \cG_{01 \, yb} 
     + \cG_{02 \, ax} \, \cG_{01 \, xy} \, \cG_{01 \, yb} \right)
\np
\sum_{xyz} 
\left( A_{t \, axb} \, \left[ B^{(124,(234,0))}_{t \, xyzb} + B^{(134,(123,0))}_{t \, xyzb} \right]  
\right.
\np
\left.
       A_{t \, azb} \, \left[ B^{(124,(234,0))}_{t \, axyz} + B^{(134,(123,0))}_{t \, axyz} \right] 
     + A_{t \, ayb} \, B^{(123,0)}_{t \, axy} \, B^{(123,0)}_{t \, yzb} \right) 
\nt 
\cG_{01 \, ax} \, \cG_{01 \, xy} \, \cG_{01 \, yz} \, \cG_{01 \, zb} \, . 
\eeq
The integration yields Eq. (\ref{G4}), which 
is written in Section~\ref{4s} using the notation 
for operators and coefficients that is described 
below in Appendix \ref{notationsummary}.

%%%%%%%%%%%%%%%%%%%%%%%%%%%%%%%%%%%%%%%%%%%%%%%%
\subsection{ Summary of terms and coefficients }
\label{notationsummary}
%%%%%%%%%%%%%%%%%%%%%%%%%%%%%%%%%%%%%%%%%%%%%%%%

The operators that emerge up to the fourth order 
in Section \ref{S} are
\beq
\label{02axb}
\cG_{02 \, axb} \es \cG_{01 \, ax} \, \cG_{01 \, xb} \, , \\
\label{03axb} 
\cG_{03 \, axb} 
\es  
\cG_{01 \, ax} \, \cG_{02 \, xb} + \cG_{02 \, ax} \, \cG_{01 \, xb} \, , \\
\label{03axyb} 
\cG_{03 \, axyb} 
\es  
\cG_{01 \, ax} \, \cG_{01 \, xy} \, \cG_{01 \, yb}
\, , \\
\label{04axb}
\cG_{04 \, axb}
\es
       \cG_{01 \, ax} \, \cG_{03 \, xb}
     + \cG_{02 \, ax} \, \cG_{02 \, xb} 
     + \cG_{03 \, ax} \, \cG_{01 \, xb} \, ,  \\
\label{04axyb}
\cG_{04 \, axyb}
\es
       \cG_{01 \, ax} \, \cG_{01 \, xy} \, \cG_{02 \, yb} 
     + \cG_{01 \, ax} \, \cG_{02 \, xy} \, \cG_{01 \, yb} 
\np    \cG_{02 \, ax} \, \cG_{01 \, xy} \, \cG_{01 \, yb} \, , \\
\label{04axyzb}
\cG_{04 \, axyzb}
\es
\cG_{01 \, ax} \, \cG_{01 \, xy} \, \cG_{01 \, yz}
\, \cG_{01 \, zb} \, .
\eeq
The corresponding coefficients are
\beq
\label{(123,0)}
B^{(123,0)}_{t \, axb}
\es
\int_0^t d\tau \, A_{\tau \, axb} \, , \\
\label{(124,(234,0))}
B^{(124,(234,0))}_{t \, axyb} 
\es
\int_0^t d\tau \, A_{\tau \, axb} \, B^{(123,0)}_{\tau \, xyb} \, , \\
\label{(134,(123,0))}
B^{(134,(123,0))}_{t \, axyb} 
\es
\int_0^t d\tau \, A_{\tau \, ayb} \, B^{(123,0)}_{\tau \, axy}  \, , \\
\label{(125,(235,(345,0)))}
B^{(125,(235,(345,0)))}_{t \, axyzb} 
\es
\int_0^t d\tau \, A_{\tau \, axb} \, B^{(124,(234,0))}_{\tau \, xyzb} \, , \\
\label{(125,(245,(234,0)))}
B^{(125,(245,(234,0)))}_{t \, axyzb} 
\es
\int_0^t d\tau \, A_{\tau \, axb} \, B^{(134,(123,0))}_{\tau \, xyzb} \, , \\
\label{(145,(124,(234,0)))}
B^{(145,(124,(234,0)))}_{t \, axyzb} 
\es
\int_0^t d\tau \, A_{\tau \, azb} \, B^{(124,(234,0))}_{\tau \, axyz}  \, , \\
\label{(145,(134,(123,0)))}
B^{(145,(134,(123,0)))}_{t \, axyzb} 
\es
\int_0^t d\tau \, A_{\tau \, azb} \, B^{(134,(123,0))}_{\tau \, axyz}  \, , \\
\label{(135,(123,0),(345,0))}
B^{(135,(123,0),(345,0))}_{t \, axyzb} 
\es
\int_0^t d\tau \, A_{\tau \, ayb} \,
B^{(123,0)}_{\tau \, axy} \, B^{(123,0)}_{\tau \, yzb} \, .
\eeq
The function $A_{t \, axb}$ is defined in Eq.
(\ref{Ataxb}). The convention for superscripts
in coefficients $B$ is designed to reflect the
origin of corresponding terms. Every coefficient 
has a subscript in the form of a list of particle 
configurations. The configurations are numbered 
with natural numbers from left to right. The 
numbers in the superscripts, indicate which 
configuration appears as a label in a corresponding 
factor under the integral. The first three numbers 
in a superscript refer to subscripts of $A$ defined
in Eq. (\ref{Ataxb}) and the remaining numbers refer 
to the subscripts in coefficients that form other 
factors under the integrals, according to the 
pattern illustrated by Eqs. (\ref{(123,0)}) 
to (\ref{(135,(123,0),(345,0))}).

\end{appendix}

%%%%%%%%%%%%%%%%%%%%%%%%%%%

%%%%%%%%%%%%%%%%%%%%%


\begin{thebibliography}{99}
%%%%%%%%%%%%%%%%%%%%%%%%%%%

\bibitem{DiracFF}
P. A. M. Dirac, 
Rev. Mod. Phys. {\bf 21}, 392 (1949).

\bibitem{GlazekWilson1}
S. D. G{\l}azek, K. G. Wilson,
Phys Rev. D {\bf 48}, 5863 (1993).

\bibitem{Wegner1}
F. Wegner, 
Ann. Phys. (Leipzig) {\bf 3}, 77 (1994).

\bibitem{RGPEP}
S. D. G{\l}azek, 
Acta Phys. Pol. B {\bf 29}, 1979 (1998).

\bibitem{Wilsonetal}
K. G. Wilson et al., 
Phys. Rev. D {\bf 49} 6720 (1994).

\bibitem{HfRGPEP}
S. D. G{\l}azek, 
%Hypothesis of quark binding by condensation of gluons in hadrons
Few-Body Syst. DOI 10.1007/s00601-011-0282-1.

\bibitem{npRGPEP}
S. D. G{\l}azek, 
Acta Phys. Pol. B {\bf 42}, 1933 (2011), 
and refs. therein.

\bibitem{Wegner2}
F. Wegner,
J. Phys. A: Math. Gen. {\bf 39}, 8221 (2006).

\bibitem{Kehrein}
S. K. Kehrein, {\it The Flow Equation Approach to Many-Particle Systems},
(Springer, 2006). 

\bibitem{Weinberg}
S. Weinberg, {\it Quantum Theory of Fields I} 
(Cambridge University Press, 1995). 

\bibitem{PDG}
K. Nakamura et al., J. Phys. G {\bf 37}, 075021 (2010).

\bibitem{partonmodel}
R. P. Feynman, 
Phys. Rev. Lett. {\bf 23}, 1415 (1969).
% R. P. Feynman, Proceedings, Conference On High Energy Collisions, 
% ed. C.N. Yang, et al, (Gordon and Breach, New York 1969), 237-258.

\bibitem{Wilson1}
K. Wilson, 
Phys. Rev. {\bf 140}, B 445 (1965).

\bibitem{Wilson2}
K. G. Wilson, 
Phys. Rev. D {\bf 2}, 1438 (1970).

\bibitem{PerryWilsonCC}
R. J. Perry, K. G. Wilson, 
Nucl. Phys. B {\bf 403}, 587 (1993).

\bibitem{WilsonFisher}
K. G. Wilson, M. E. Fisher, 
Phys. Rev. Lett. {\bf 28}, 240 (1972).

\bibitem{lc1}
K. G. Wilson
%Renormalization Group and Strong Interactions 
Phys. Rev. D {\bf 3}, 1818 (1971).

\bibitem{lc2}
S. D. G{\l}azek, K. G. Wilson
Phys. Rev. B {\bf 69}, 094304 (2004). 

\bibitem{lcet}
S. D. G{\l}azek
%Limit cycles of effective theories 
Phys. Rev. D {\bf 75}, 025005 (2007). 

\bibitem{GlazekScalar}
S. D. G{\l}azek, 
Phys. Rev. D {\bf 60}, 105030 (1999).

\bibitem{gluons}
S. D. G{\l}azek, 
Phys. Rev. D {\bf 63}, 116006 (2001). 

\bibitem{DiracDeadWood}
P. A. M. Dirac, 
Phys. Rev. {\bf 139}, B 684 (1965).

\bibitem{AlteredWegner}  
S. D. G{\l}azek, J. M{\l}ynik,
%Optimization of perturbative similarity renormalization 
%group for Hamiltonians with asymptotic freedom and bound states
Phys. Rev. D {\bf 67}, 045001 (2003).

\bibitem{AccuracyEstimate}
S. D. G{\l}azek, J. M{\l}ynik,
%Accuracy estimate for a relativistic Hamiltonian approach 
%to bound state problems in theories with asymptotic freedom
Acta Phys. Pol. B {\bf 35}, 723 (2004). 

\bibitem{HQ}
S. D. G{\l}azek, J. M{\l}ynik, 
Phys. Rev. D {\bf 74}, 105015 (2006).

%%%%%%%%%%%%%%%%%%%%%
\end{thebibliography}
\end{document}